\documentclass[pra,twocolumn,a4paper,groupedaddress,showpacs,floatfix,aps,10pt]{revtex4-1}
\usepackage{graphicx,amsmath,amssymb,amsfonts,dsfont}
\newcommand{\mf}[1]{\boldsymbol{#1}}
\newcommand{\ket}[1]{\ensuremath{|#1\rangle}}
\newcommand{\mc}[1]{\ensuremath{\mathcal{#1}}}
\newcommand{\bra}[1]{\ensuremath{\langle #1 |}}

\usepackage{color}

\newcommand{\vroS}{R}
\newcommand{\vroW}{\varrho}

\newcommand{\imag}{\mathrm{i}}
\begin{document}

\title{Two-way interconversion of millimeter-wave and optical fields in Rydberg gases}

\author{Martin Kiffner${}^{1,2}$}
\author{Amir Feizpour${}^{2}$}
\author{Krzysztof T.~Kaczmarek${}^{2}$}
\author{Dieter Jaksch${}^{2,1}$}
\author{Joshua Nunn${}^{2}$}

\affiliation{Centre for Quantum Technologies, National University of Singapore,
3 Science Drive 2, Singapore 117543${}^1$}
\affiliation{Clarendon Laboratory, University of Oxford, Parks Road, Oxford OX1
3PU, United Kingdom${}^2$}

%
%\pacs{42.50.Gy,42.65.Ky,32.80.Ee}
%
% 42.50.Gy Effects of atomic coherence on propagation, absorption, and amplification of light; electromagnetically induced transparency and absorption
%
% 42.65.Ky Frequency conversion; harmonic generation, including higher-order harmonic generation
%
% 32.80.Ee Rydberg states
%

%
\begin{abstract}
We show that cold Rydberg gases enable an efficient six-wave mixing process where terahertz or 
microwave fields are coherently converted into optical fields and vice versa. This process 
is made possible by the long lifetime of Rydberg states, the strong coupling of millimeter 
waves to Rydberg transitions and by a quantum interference effect related to electromagnetically 
induced transparency (EIT). 
Our frequency conversion scheme applies to a broad 
spectrum of millimeter waves due to the abundance of transitions within the Rydberg manifold, 
and we discuss two possible  implementations  based on focussed terahertz 
beams and millimeter wave fields confined by a waveguide, respectively.
We analyse a realistic example for  the interconversion of terahertz and optical fields 
in rubidium atoms and find that the conversion efficiency can in principle exceed 90\%. 
\end{abstract}

\maketitle

\section{Introduction \label{intro}}
Two-way conversion between optical fields and terahertz/microwave radiation is a highly 
desirable capability with applications in classical and quantum technologies, including the metrological transfer of atomic frequency standards~\cite{Fortier:2011aa}, 
novel astronomical surveys~\cite{Martin:2012aa}, 
long-distance transmission of electronic data via photonic carriers~\cite{Yang:2014aa}, and signal processing for 
applications in radar and avionics~\cite{Loic:2009aa}. Efficient conversion of terahertz radiation into visible light would facilitate the generation, 
detection and imaging of terahertz fields~\cite{adam:11,chan:07} for stand-off detection, biomedical diagnostics and spectroscopy.
In the quantum domain, coherent microwave-optical conversion could enable quantum computing via optically-mediated entanglement 
swapping~\cite{Barrett:2005aa,Monroe:2014aa,Nemoto:2014aa} in solid state systems
such as spins in silicon~\cite{Morton:2015aa} 
or superconducting qubits~\cite{Barends:2014aa}, which lack optical transitions but couple strongly to microwaves.
Moreover, Josephson junctions can mediate microwave photonic non-linearities that cannot easily be replicated for optical photons~\cite{Wallraff:2004aa} 
so that coherent microwave-optical conversion also provides a route to freely-scalable all-photonic quantum computing.
Recent proposals for conversion between the optical and mm-wave domains have been based on optomechanical transduction~\cite{Andrews:2014aa,Bagci:2014aa,xia:14}, or frequency 
mixing in $\Lambda$-type atomic ensembles~\cite{Williamson:2014aa,OBrien:2014aa,Blum:2015aa,Hafezi:2012aa,Marcos:2010aa}. 
Both approaches require high quality frequency-selective cavities limiting 
the conversion bandwidth, as well as aggressive cooling or optical pumping 
to bring the conversion devices into their quantum ground states. 

\begin{figure*}[t!]
\begin{center}
\includegraphics[width=12cm]{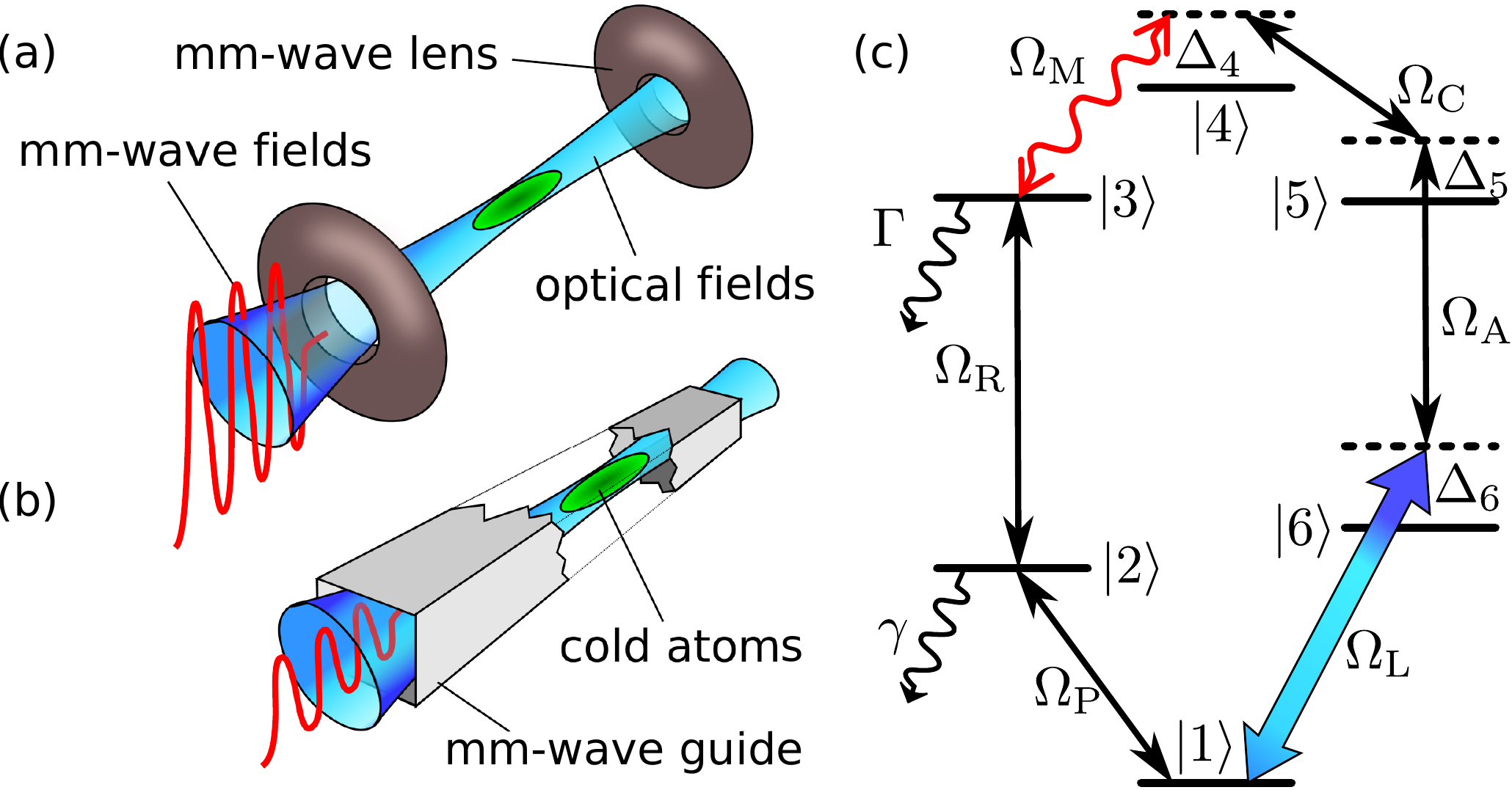}
\end{center}
\caption{\label{fig1}
(a) and (b) depict two possible implementations of the considered MMOC device where 
a mm-wave field (red) and an optical field (blue) are coherently interconverted through their interaction with an ensemble of cold atoms (green). The other fields needed to drive the interaction are not shown. 
(a) The mm-wave fields are coupled into and out of the atomic ensemble with dielectric lenses; the optical fields are directed through small, sub-mm holes. 
(b) Tight confinement of microwaves with longer wavelengths could be achieved by trapping the atoms inside the core of a hollow waveguide.
(c) Atomic level scheme.
Transition frequencies and detunings are not 
to scale. $\Omega_{\text{M}}$ and $\Omega_{\text{L}}$ are the 
Rabi frequencies associated with the mm-wave and the optical fields, 
respectively. 
 $\Omega_{\text{P}}$, $\Omega_{\text{R}}$, $\Omega_{\text{C}}$ and 
$\Omega_{\text{A}}$ are Rabi frequencies of the auxiliary fields, 
 and $\Delta_k$  is the detuning of the fields with state $\ket{k}$ 
($k\in\{4,5,6\})$. Levels $\ket{3}$, $\ket{4}$ and $\ket{5}$ are Rydberg 
states with decay rate $\Gamma\ll\gamma$, where $\gamma$ is the decay rate of states $\ket{2}$ and $\ket{6}$. 
}
\end{figure*}

In this paper, we propose instead to use frequency mixing in Rydberg gases~\cite{huber:14,che:15,zhang:15} for the conversion of millimeter waves to optical fields (MMOC) (see~Fig.~\ref{fig1}). 
We use the terminology `mm-wave' broadly to refer to fields with carrier frequencies between $10$ and $10,000$~GHz, 
corresponding to resonant transitions between highly excited Rydberg states in an atomic vapour.
Our scheme benefits from the strong coupling between Rydberg atoms and millimeter waves which has previously been used for detection and magnetometry~\cite{sedlacek:12,gordon:14,fan:15}, 
storage of microwaves~\cite{petrosyan:09} and hybrid atom-photon gates~\cite{pritchard:14}. 
Here we show how to achieve efficient and coherent MMOC without the need for cavities, microfabrication or cooling. 
Our MMOC scheme is made possible by an EIT-related quantum interference effect and the long lifetime of the Rydberg states. 
In contrast to previous frequency mixing schemes in EIT media~\cite{fleischhauer:05,zhang:09,zhang:11}, this quantum interference 
effect implements a coherent beam splitter interaction between the millimeter and optical fields which underpins the conversion effect. 
Our main result is a theoretical model establishing the principle of operation of the proposed device, which it is 
shown could be implemented in an ensemble of cold trapped Rb atoms.
The paper is organised as follows. We introduce our theoretical model based on the standard 
framework of coupled Maxwell-Bloch equations in Sec.~\ref{methods}, where we also 
describe how to include interactions between Rydberg atoms. 
In Sec.~\ref{results} we discuss the principle of operation 
of our scheme and show that both time-independent and pulsed input fields of arbitrary (band-limited) 
shape can be efficiently converted. 
We go on to consider the simultaneous spatial confinement of mm-wave and optical fields, and 
we show that high conversion efficiencies are predicted for a realistic implementation in trapped Rb vapour. 
A brief summary of our work is presented in Sec.~\ref{summary}. 
\section{Model \label{methods}}
%
%
%%%%%%%%%%%%%%%%%%%%%%
%%%%%%%%%%%%%%%%%%%% 
%
We consider an ensemble of cold trapped atoms interacting with laser fields and mm-waves 
and model these interactions using the standard framework of coupled Maxwell-Bloch equations. 
A  summary  of the general approach is presented in   Sec.~\ref{maxbloch}, 
and a detailed derivation can be found in the Supplementary Information. 
The analytical solution of the Maxwell-Bloch equations is outlined in Sec.~\ref{ana} and complemented by Appendix~\ref{solution}.  
In Sec.~\ref{imperfections} we include  Rydberg-Rydberg interactions into our model. 
This allows us to identify parameter regimes in Sec.~\ref{results} where these interactions are negligible. 
\subsection{Maxwell-Bloch equations \label{maxbloch}} 
In a first step we neglect atom-atom interactions and  consider the Bloch equations for a single atom with level scheme as shown in Fig.~\ref{fig1}. 
The millimeter wave $\Omega_{\text{M}}$ of interest couples to the transition $\ket{3}\leftrightarrow\ket{4}$, 
where  $\ket{3}$ and $\ket{4}$ are Rydberg states with  principal quantum number $n\gtrsim 20$.  
The optical field $\Omega_{\text{L}}$ of interest couples to the 
$\ket{1}\leftrightarrow\ket{6}$ transition, and the conversion between $\Omega_{\text{M}}$ and $\Omega_{\text{L}}$ is 
facilitated by four auxiliary fields. 
The resonant fields $\Omega_{\text{P}}$ and $\Omega_{\text{R}}$ create a coherence on the 
$\ket{1}\leftrightarrow\ket{3}$ transition through coherent population trapping~\cite{arimondo:cpt}. 
The two other auxiliary fields $\Omega_{\text{C}}$ and $\Omega_{\text{A}}$ are in general off-resonant and establish a coherent connection 
between the $\ket{3}\leftrightarrow\ket{4}$ and $\ket{1}\leftrightarrow\ket{6}$ transitions. 
We model the time evolution of the atomic density operator by a Markovian master equation
\begin{align}
\partial_t \vroW &= - \frac{\imag}{\hbar} [ H , \vroW ] +\mc{L}_{\gamma}\vroW\,. 
\label{master_eq}
\end{align} 
In the electric-dipole and  rotating-wave approximations, the  
Hamiltonian $H$ in Eq.~(\ref{master_eq}) is given by 
\begin{align}
H  = & - \hbar\sum\limits_{k=3}^{6} \Delta_{k} A_{kk}   
 -\hbar \left( \Omega_{\text{P}}A_{21}+ \Omega_{\text{R}}A_{32}+\Omega_{\text{M}}A_{43} \right.\notag \\
&\qquad \left. + \Omega_{\text{C}}A_{45}+ \Omega_{\text{A}}A_{56} + \Omega_{\text{L}}A_{61}  \,+\,\text{H.c.}\right)\, ,
\label{H}
\end{align}
and  
$A_{ij} = \ket{i}\bra{j}$ 
are atomic transition operators. The detuning $\Delta_{k}$ in Eq.~(\ref{H}) is defined as 
 \begin{subequations}
 \label{detuning}
 \begin{align} 
\Delta_{3}=& \omega_{\text{P}}+\omega_{\text{R}} -\omega_{3}\,, \\
\Delta_{4}=& \omega_{\text{P}}+\omega_{\text{R}}+\omega_{\text{M}} -\omega_{4}\,, \\
\Delta_{5}=& \omega_{\text{P}}+\omega_{\text{R}}+\omega_{\text{M}} - \omega_{\text{C}} -\omega_{5}\,, \\
\Delta_{6}=& \omega_{\text{L}} -\omega_{6}\,,
\end{align}
\end{subequations}
where $\hbar\omega_{k}$ denotes the energy of state $\ket{k}$ with respect to the energy of level $\ket{1}$ and $\omega_{\text{X}}$ 
is the frequency of field X with Rabi frequency   $\Omega_{\text{X}}$  ($\text{X}\in\{\text{P},\text{R},\text{C},\text{A},\text{M},\text{L}\}$). 
The term $\mc{L}_{\gamma}\vroW$ in Eq.~(\ref{master_eq}) accounts for spontaneous emission from the excited states. These processes are described 
by standard Lindblad decay terms. The full decay rate of the states 
$\ket{2}$ and $\ket{6}$ is $\gamma$, and the long-lived Rydberg states decay with $\Gamma\ll\gamma$. 
The six fields  drive a resonant loop,
\begin{align}
 \omega_{\text{P}}+\omega_{\text{R}}+\omega_{\text{M}}-\omega_{\text{C}}-\omega_{\text{A}}-\omega_{\text{L}}=0\,, 
 \label{freq_cond}
\end{align}
and we impose the phase matching condition 
\begin{align}
 \mf{k}_{\text{P}}+\mf{k}_{\text{R}}+\mf{k}_{\text{M}}-\mf{k}_{\text{C}}-\mf{k}_{\text{A}}=\mf{k}_{\text{L}}\,.
 \label{phase_matching}
\end{align}
In the following, we assume that 
$\Omega_{\text{M}}$ and $\Omega_{\text{L}}$ are co-propagating, while the directions of the auxiliary fields are chosen such that Eq.~(\ref{phase_matching}) holds. Note 
that this phase matching condition is automatically fulfilled by virtue of Eq.~(\ref{freq_cond}) if all fields are co-propagating.
The strong auxiliary fields are not significantly affected by their interaction with the mm-wave and optical signals. 
We therefore consider only these signal fields and the atomic coherences as dynamical variables.
In the paraxial approximation we find 
\begin{subequations}
\label{maxS1d} 
\begin{align}
& \left(-\frac{\imag}{2k_{\text{M}}}\Delta_{\perp}+ \frac{1}{c} \partial_{t} + \partial_{z} \right)  \Omega_{\text{M}} = \imag  \eta_{\text{M}} \vroW_{43} \,, \label{maxS11d} \\
& \left(-\frac{\imag}{2k_{\text{L}}}\Delta_{\perp}+\frac{1}{c} \partial_{t}+ \partial_{z}\right)  \Omega_{\text{L}}  =  \imag \eta_{\text{L}} \vroW_{61}\, ,  \label{maxS21d} 
 \end{align}
\end{subequations}
where $k_{\text{M}}$ $(k_{\text{L}})$ is the wavenumber of the mm-wave (optical) field and $\Delta_{\perp}=\partial_x^2+\partial_y^2$ is the transverse Laplace operator. 
The coupling constants $\eta_{\text{M}}$ and $\eta_{\text{L}}$ are given by 
 \begin{subequations}
 \begin{align}
 \eta_{\text{M}} & =\mc{N} \frac{|\mf{d}_{43}|^2}{2\hbar\epsilon_0 c} \omega_{\text{M}} \,, \\
 \eta_{\text{L}} & =\mc{N} \frac{ |\mf{d}_{61}|^2}{2\hbar\epsilon_0 c} \omega_{\text{L}} \,, 
 \end{align}
 \end{subequations}
 where 
$\mf{d}_{kl}=\bra{k}\mf{\hat{d}}\ket{l}$
is the  matrix element of the electric dipole moment operator $\mf{\hat{d}}$ on the transition transition $\ket{k}\leftrightarrow\ket{l}$, 
$c$ is the  speed of light and $\mc{N}$ is the density of atoms. 
In the following the ratio of the coupling constants is denoted by 
 \begin{align}
  \label{bval}
  b^2=\eta_{\text{M}}/\eta_{\text{L}} = \frac{|\mf{d}_{43}|^2}{|\mf{d}_{61}|^2} \frac{\omega_{\text{M}}}{\omega_{\text{L}}}.
 \end{align}
Numerical solutions of Eqs.~(\ref{master_eq}),~(\ref{maxS1d}) are presented in Sec.~\ref{imp}, 
but first it is instructive to derive analytic solutions in the limit that diffraction over the length of the atomic ensemble can be neglected.
\subsection{Analytical solution \label{ana}}
The first-order solution of Eq.~(\ref{master_eq}) with respect to the Rabi frequencies $\Omega_{\text{M}}$, $\Omega_{\text{L}}$ 
takes the form (see Appendix~\ref{solution})
\begin{subequations}
  \label{psol}
\begin{align}
 \vroW_{43} \approx  \;& \chi_{43}^{\text{M}} \Omega_{\text{M}} + \chi_{43}^{\text{L}}\Omega_{\text{L}} \,,  \\
 \vroW_{61} \approx  \;& \chi_{61}^{\text{M}}\Omega_{\text{M}} +\chi_{61}^{\text{L}} \Omega_{\text{L}} \,. 
\end{align}
\end{subequations}
The response of the atomic system on the $\ket{3}\leftrightarrow\ket{4}$ transition induced by the mm-wave field is described by $\chi_{43}^{\text{M}}$, and 
$\chi_{61}^{\text{L}}$ accounts for the atomic response on the transition $\ket{1}\leftrightarrow\ket{6}$ due to the optical field.  In addition, 
the mm-wave field can induce a coherence proportional to $\chi_{61}^{\text{M}}$ on the optical transition $\ket{1}\leftrightarrow\ket{6}$, 
and the optical field can create a coherence proportional to $\chi_{43}^{\text{L}}$ on the transition $\ket{3}\leftrightarrow\ket{4}$. 
The cross-terms proportional to $\chi_{43}^{\text{L}}$  and $\chi_{61}^{\text{M}}$  in Eq.~(\ref{psol}) originate from the closed-loop character of the atomic level scheme.

Next we combine Eq.~(\ref{psol}) with Eq.~(\ref{maxS1d}) and make the simplifying assumption that diffraction over the ensemble length can be neglected, so that the transverse Laplacians can be dropped. 
Making a coordinate transformation from the laboratory frame $(t,z)$ to a frame $(\tau = t-z/c, z)$ co-moving with the signal fields, 
the evolution equation for the mm-wave and optical fields can then be written as 
\begin{align}
\partial_{z}\mf{\Omega} = \imag \mc{M} \mf{\Omega},
 \label{mateq}
\end{align}
where 
\begin{align}
\mc{M}=  \eta_{\text{L}}\left(
 \begin{array}{cc}
  b^2 \chi_{43}^{\text{M}} &  b^2 \chi_{43}^{\text{L}} \\[0.2cm]
 \chi_{61}^{\text{M}} & \chi_{61}^{\text{L}} 
 \end{array}
\right), &\quad 
 \mf{\Omega}=\left( 
 \begin{array}{c} 
 \Omega_{\text{M}} \\ 
 \Omega_{\text{L}} 
 \end{array} 
 \right)\,. 
 \label{matrixM}
\end{align}
When the auxiliary fields are time-independent and spatially uniform, the solution to Eq.~(\ref{mateq}) is
\begin{align}
 \mf{\Omega}=\exp(\imag \mc{M} z) \mf{\Omega}_0,
 \label{gensol}
\end{align}
where $\mf{\Omega}_0$ is the initial condition $\mf{\Omega}$ evaluated at $z=0$. The matrix exponential in Eq.~(\ref{gensol}) can be 
expressed in terms of the $2\times2$ identity matrix $\mathds{1}$ and the Pauli matrices $\sigma_k$~\cite{tannoudji:qm}, 
\begin{align}
 \exp(\imag \mc{M} z)= &\exp(\imag a_0 z)\notag \\
 &\left[ \cos\left(\sqrt{\mf{a}^2} z\right)\mathds{1} + \imag \frac{\mf{a}\cdot\mf{\sigma}}{\sqrt{\mf{a}^2}}\sin\left(\sqrt{\mf{a}^2}z\right) \right]\,,
 \label{generalexp}
\end{align}
where  
\begin{align}
 a_0=\frac{1}{2}\text{Tr}(\mc{M})\,,\quad \mf{a} = \frac{1}{2}\text{Tr}(\mc{M}\mf{\sigma})\,.
 \label{acoff}
\end{align}
The solution presented here treats the signal fields $\Omega_{\text{M}}$, $\Omega_{\text{L}}$ as $c$-numbers. However, the generalisation 
to quantum fields is straightforward since the coherences in Eq.~(\ref{psol}) are linear in the signal fields. Apart from quantum noise operators, our  calculations are 
thus equivalent to a Heisenberg-Langevin approach where the signal Rabi frequencies $\Omega_{\text{M}}$, $\Omega_{\text{L}}$ are replaced 
by quantum fields~\cite{fleischhauer:02,hafezi:09,zimmer:08}. Since the Langevin noise operators  
represent only vacuum noise, they do not contribute to  normally ordered expectation values, which determine the conversion efficiency.
%
%%%%%%%%%%%%%%%%%%%%%%
\begin{figure*}[t!]
\begin{center}
\includegraphics[width=15cm]{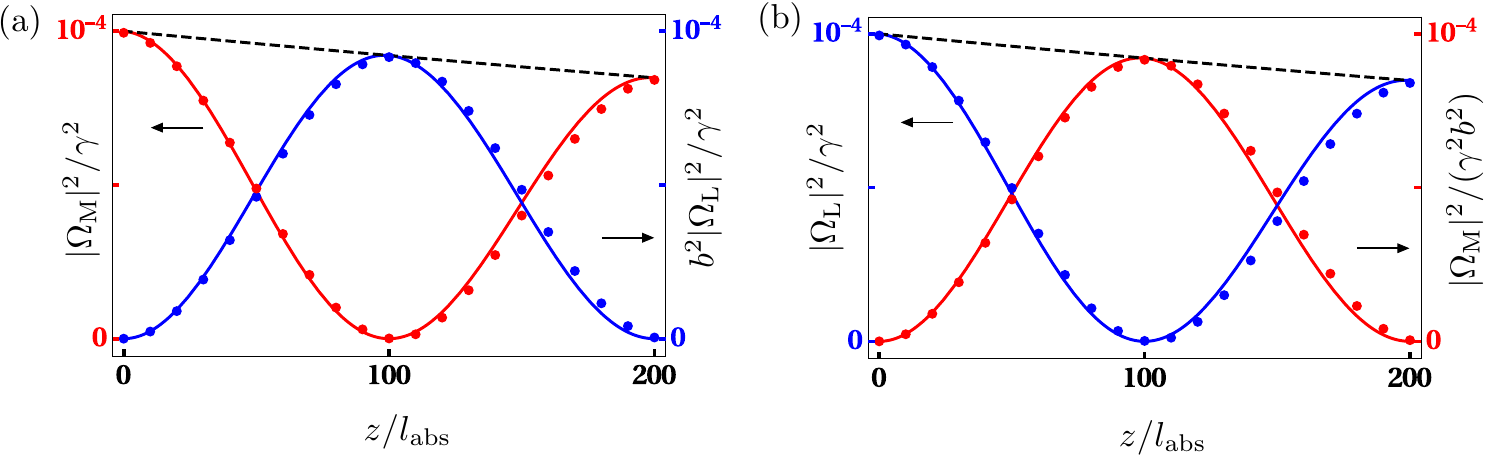}
\end{center}
\caption{\label{fig2}
Frequency conversion of stationary fields.  Intensities of the 
millimeter wave (red) and optical field (blue)  inside the medium. 
Dots indicate the results from a numerical integration of Maxwell-Bloch 
equations. 
(a) A CW millimeter wave enters the medium at $z=0$. 
(b) A CW optical field enters the medium at $z=0$. 
In (a) and (b), the dashed line is proportional to the envelope 
$e^{-2 \kappa z}$ and  we set $\Gamma/\gamma= 1/285$,  
$\Omega_{\text{A}}=2 \gamma$, $\Omega_{\text{C}}=2 \gamma$, 
 $\Omega_{\text{R}}=2 \gamma$, $\Omega_{\text{P}}=0.3 \gamma$, $\Delta_{4}=2 
\gamma$, $\Delta_{5}=2 \gamma$, $\Delta_{6}=2 \gamma$ and $b=\sqrt{0.72}$. These parameters correspond to the realisation of 
the  level scheme with Rubidium atoms in Sec.~\ref{real}.
}
\end{figure*}
%%%%%%%%%%%%%%%%%%%% 
%
\subsection{Interaction-induced imperfections \label{imperfections}}
Next we consider the effects of dipole-mediated interactions between atoms excited into their Rydberg manifolds. 
In general, Rydberg interactions will prevent some fraction of atoms from participating in the conversion process and lead to absorption of 
the signal fields, and therefore will reduce the conversion efficiency. 
The atomic level scheme in Fig.~\ref{fig1}(b) 
contains three Rydberg states, and the population in state $\ket{3}$ is continuously kept at 
$\rho_{33}\approx|\Omega_{\text{P}}/\Omega_{\text{R}}|^2$ via coherent population trapping. 
On the other hand, the population in the other Rydberg states $\ket{4}$ and $\ket{5}$ is negligibly small for 
weak fields $\Omega_{\text{M}}$ and $\Omega_{\text{L}}$. The dominant perturbation to the conversion mechanism will thus stem from nearby Rydberg atoms 
in state $\ket{3}$. In order to model this, we consider a system of two atoms 
where  atom A is located at the coordinate origin. The conversion process in atom A is disturbed by Rydberg-Rydberg interactions with atom B, which is 
prepared in state $\ket{3}$ and  positioned at  $\mf{R}$. 
Next we discuss the two dominant effects caused by the presence of atom B. 
First, atom B gives rise to a van der Waals shift of state $\ket{3}$ in atom A~\cite{singer:05}, 
\begin{align}
  \hbar \Delta_{\text{vdW}} = -\frac{C_6}{R^6} ,
 \label{vdW}
\end{align}
where the coefficient $C_6$ depends on the quantum numbers of state $\ket{3}$. 
If $R$ is smaller than the blockade radius $R_{\text{b}}$, atom A cannot be excited to the Rydberg state and thus does not participate in the conversion.  
The blockade radius is determined by the single-atom EIT linewidth $\gamma_{\text{EIT}}=|\Omega_{\text{R}}|^2/\gamma$ 
and given by $R_{\text{b}}=[2 |C_6|/(\hbar\gamma_{\text{EIT}})]^{1/6}$~\cite{peyronnel:12}. 
Second, atom B gives rise to a frequency shift of state $\ket{4}$ in atom A via the resonant dipole-dipole interaction~\cite{gallagher:ryd}, 
\begin{align}
\hbar \Delta_{\text{DD}} = \frac{1}{4\pi\varepsilon_0}\frac{|\mf{d}_{43}|^2-3|\mf{d}_{43}\cdot\vec{\mf{R}}|^2}{R^3},
\label{dd}
\end{align}
where  $\vec{\mf{R}}=\mf{R}/R$. In contrast to the 
van der Waals shift in Eq.~(\ref{vdW}), $\Delta_{\text{DD}}$ depends on the relative orientation of the two atoms. 
In principle, state $\ket{5}$ in atom A can experience a similar shift $\Delta_{\text{DD}}$ if the dipole moment $\mf{d}_{53}$ is 
different from zero. Here we assume that states $\ket{5}$ and $\ket{3}$ have the same parity so that $\mf{d}_{53}=0$, consistent with the example implementation in Rb that we introduce below in Sec.~\ref{real}.
The preceding discussion shows that Rydberg-Rydberg interactions change the energies of states $\ket{3}$ and $\ket{4}$. 
In order to incorporate these frequency shifts into our model, we find the general first-order solution of the atomic coherences in Eq.~(\ref{psol}) 
for arbitrary detunings and Rabi frequencies of the auxiliary fields. We then introduce the 
effective detuning parameters
\begin{subequations}
\begin{align}
\widetilde{\Delta}_3 & = \Delta_3 - \Delta_{\text{vdW}} , \\
\widetilde{\Delta}_4 & = \Delta_4 - \Delta_{\text{DD}},
\end{align}
\end{subequations}
and replace $\Delta_3$ and $\Delta_4$ in the general expression for the matrix $\mc{M}$ in Eq.~(\ref{matrixM}) by $\widetilde{\Delta}_3$ and $\widetilde{\Delta}_4$. 
Since $\Delta_{\text{vdW}}$ and $\Delta_{\text{DD}}$ depend on the relative position $\mf{R}$, we average $\mc{M}$ over $\mf{R}$,
\begin{align}
\mc{M} \longrightarrow \widetilde{\mc{M}} = \int \!\!\text{d}^3\!\mf{R}\;\, \mc{M}(\mf{R})w(\mf{R})\,,
 \label{averM}
\end{align}
where the distribution of nearest neighbours in a random sample of Rydberg atoms follows the probability density~\cite{chandrasekhar:43}, 
\begin{align}
w(\mf{R}) = \frac{1}{4\pi} \frac{3}{r_{\text{ws}}}\left(\frac{R}{r_{\text{ws}}}\right)^2 \exp\left[-\left(\frac{R}{r_{\text{ws}}} \right)^3\right]\,,
\label{distribution}
\end{align}
with the parameter
\begin{align}
 r_{\text{ws}}= \left[\frac{3}{4\pi \mc{N}_{\text{Ry}}}\right]^{1/3}
\end{align}
the Wigner-Seitz radius for a given density of Rydberg atoms $\mc{N}_{\text{Ry}}$. 

This account of Rydberg-Rydberg interactions is expected to work well for weak optical and mm-wave fields. If 
the intensities of $\Omega_{\text{M}}$ and $\Omega_{\text{L}}$ are increased such that the population in $\ket{4}$ and $\ket{5}$ is not 
negligible, other dipole-dipole interactions can occur that are not captured by our model. 
Furthermore, our model neglects cooperative effects like superradiance~\cite{wang:07} 
and frequency shifts due to a ground state atom within the electron orbit of a Rydberg state~\cite{bendkowsky:09}. 
However, experimental results~\cite{weatherill:08,han:15,pritchard:ar} for EIT involving a Rydberg state show that these effects 
can be negligible for low principal quantum numbers $n\lesssim 40$, for weak probe fields and low atomic densities. 
\section{Results \label{results}}
In a first step we analyse the simplified analytical model of Sec.~\ref{ana} in order 
to explain the principle of the conversion mechanism. This is presented  in Sec.~\ref{ideal} 
where we also investigate the maximally achievable conversion efficiencies. 
We then introduce one possible implementation of our scheme in rubidium vapour in Sec.~\ref{real} and find a set 
of parameters for which Rydberg-Rydberg interactions are negligibly small. 
Finally, we present numerical results for MMOC in the physical 
systems shown in Figs.~\ref{fig1} (a) and (b) in Sec.~\ref{imp}. 
\subsection{Conversion mechanism\label{ideal}}
The conversion efficiency between mm-wave and optical fields according to Eq.~(\ref{gensol}) will be small for a generic matrix $\mc{M}$, but complete conversion 
can be achieved if the atomic ensemble realises a beam splitter interaction
\begin{align}
V_{\text{BS}} \propto \left(\hat{\Omega}_{\text{M}}\hat{\Omega}_{\text{L}}^{\dagger} + \hat{\Omega}_{\text{M}}^{\dagger}\hat{\Omega}_{\text{L}} \right)\,,
\end{align}
where the `hat' notation emphasises the operator nature of the fields.  
Formally, such an interaction corresponds to the case where the diagonal elements of $\mc{M}$ vanish. 
We find that this condition, such that $\chi_{43}^{\text{M}}\approx\chi_{61}^{\text{L}}\approx 0$, can indeed be met if the intensities and detunings of the auxiliary fields satisfy
\begin{align}
 |\Omega_{\text{R}}|\gg|\Omega_{\text{P}}|\,,\quad \Delta_5  =\frac{|\Omega_{\text{C}}|^2}{\Delta_4} \,, \quad \Delta_6  
=\frac{|\Omega_{\text{A}}|^2}{\Delta_5}\,. 
 \label{cond}
\end{align}
To first order in $\Gamma/\gamma$ the susceptibilities in Eq.~(\ref{psol}) are then given by 
\begin{subequations}
\label{psol2}
\begin{align}
 \chi_{43}^{\text{M}} \approx & \imag \frac{8}{b^2 \gamma} \varepsilon^2\,,\quad \chi_{43}^{\text{L}} \approx \alpha \,,  \\
 \chi_{61}^{\text{M}} \approx & \alpha^* \,, \quad \chi_{61}^{\text{L}} \approx \imag \frac{8}{\gamma} \varepsilon_{\Gamma} \,,
\end{align}
\end{subequations}
where 
\begin{subequations}
 \label{pcoeff}
\begin{align}
 \alpha = & -\frac{\Omega_{\text{C}} \Omega_{\text{P}}^*}{\Delta_4 \Omega_{\text{A}}^{*}\Omega_{\text{R}}}   \,,  \label{alphaS} \\
  \varepsilon= & \frac{b}{4} \frac{ 
\gamma}{|\Delta_4|}\frac{|\Omega_{\text{C}}|}{|\Omega_{\text{A}}|}\frac{|\Omega_
{\text{P}}|}{|\Omega_{\text{R}}|}  \,,\quad  \label{eps} \\
\varepsilon_{\Gamma}= & \frac{\Gamma \gamma}{16|\Omega_{\text{A}}|^2} 
\left(1+2\frac{|\Omega_{\text{C}}|^2}{\Delta_{4}^2}\right)\,.  \label{epsgamma}
\end{align}
\end{subequations}
$\varepsilon$ and $\varepsilon_{\Gamma}$ are dimensionless parameters that are generally smaller than unity. Since $\varepsilon_{\Gamma}\propto\Gamma$, $\varepsilon_{\Gamma}$ 
is typically of the order of $\varepsilon^2$. On the other hand, $|\alpha|\propto \varepsilon$ and hence the off-diagonal elements of the matrix $\mc{M}$ are 
indeed much larger than the diagonal elements. 
%
%%%%%%%%%%%%%%%%%%%%%%
\begin{figure}[t!]
\begin{center}
\includegraphics[width=7.5cm]{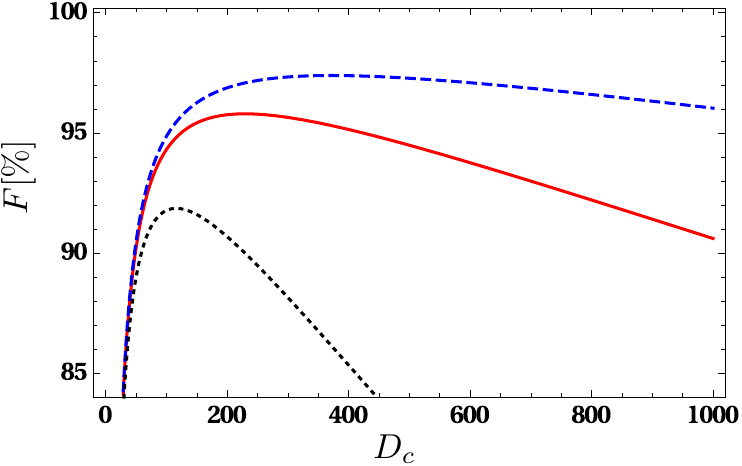}
\end{center}
\caption{\label{fig3}
Conversion efficiency as a function of optical depth $D_c$. We set $\Gamma/\gamma=3.9\times 10^{-3}$ (black dotted line), 
$\Gamma/\gamma= 10^{-3}$ (red solid line) and 
$\Gamma/\gamma= 3.8\times10^{-4}$ (blue dashed line). These parameters correspond 
to rubidium Rydberg states at zero temperature with $n\approx 20$, $n\approx 30$ and $n\approx 40$, respectively.
Common parameters in all three curves are $\Omega_{\text{A}}=2 \gamma$, 
$\Omega_{\text{C}}=2 \gamma$ and $\Delta_{4}=2 \gamma$. 
}
\end{figure}
%%%%%%%%%%%%%%%%%%%% 
%

%
This result can be understood as follows. The level scheme in Fig.~\ref{fig1} can be regarded as three consecutive EIT systems where the  weak probe fields are represented by 
$\Omega_{\text{P}}$, $\Omega_{\text{M}}$ and $\Omega_{\text{L}}$, respectively. However, these three systems are coupled and hence the normal two-photon resonance condition 
for transparency of the $\Omega_{\text{M}}$ and $\Omega_{\text{L}}$ fields is changed.  
The conditions in Eq.~(\ref{cond}) approximately restore transparency for the field  $\Omega_{\text{M}}$ ($\Omega_{\text{L}}$) 
on the transition $\ket{3}\leftrightarrow\ket{4}$ ($\ket{1}\leftrightarrow\ket{6}$) and in the presence of the other levels and fields such that $\chi_{43}^{\text{M}}\approx 0$ 
($\chi_{61}^{\text{L}}\approx 0$). However, $\Omega_{\text{M}}$ still creates 
a coherence on the optical transition and $\Omega_{\text{L}}$ induces a coherence on the Rydberg transition such that the fields are interconverted as they propagate along the medium. 
With Eqs.~(\ref{psol2}) and~(\ref{pcoeff}) the general solution for the spatial distribution of the fields in Eq.~(\ref{gensol}) is given by  
\begin{align}
\mf{\Omega}(z) \approx
e^{-\kappa z}
\left(
 \begin{array}{cc}
  \cos(k z) & \imag b \sin(k z) \\
  \frac{\imag}{b} \sin(k z) & \cos(k  z)
 \end{array}
\right)
\mf{\Omega}(0)\, , 
 \label{matrix}
\end{align}
where 
 $\kappa=(\varepsilon^2+\varepsilon_{\Gamma})/l_{\text{abs}}$ and  
 $k=\varepsilon/l_{\text{abs}}$ determine the loss and 
 the spatial oscillation period of the interconversion, respectively, 
$l_{\text{abs}} = \gamma/(4\eta_{\text{L}})$ is the resonant absorption length 
on the $\ket{6}\leftrightarrow\ket{1}$ transition 
and  we assumed $\varepsilon\ll 1$ and $\varepsilon_{\Gamma}/\varepsilon\ll 1$. 
%
%%%%%%%%%%%%%%%%%%%%%%
\begin{figure*}[t!]
\begin{center}
\includegraphics[height=5.5cm]{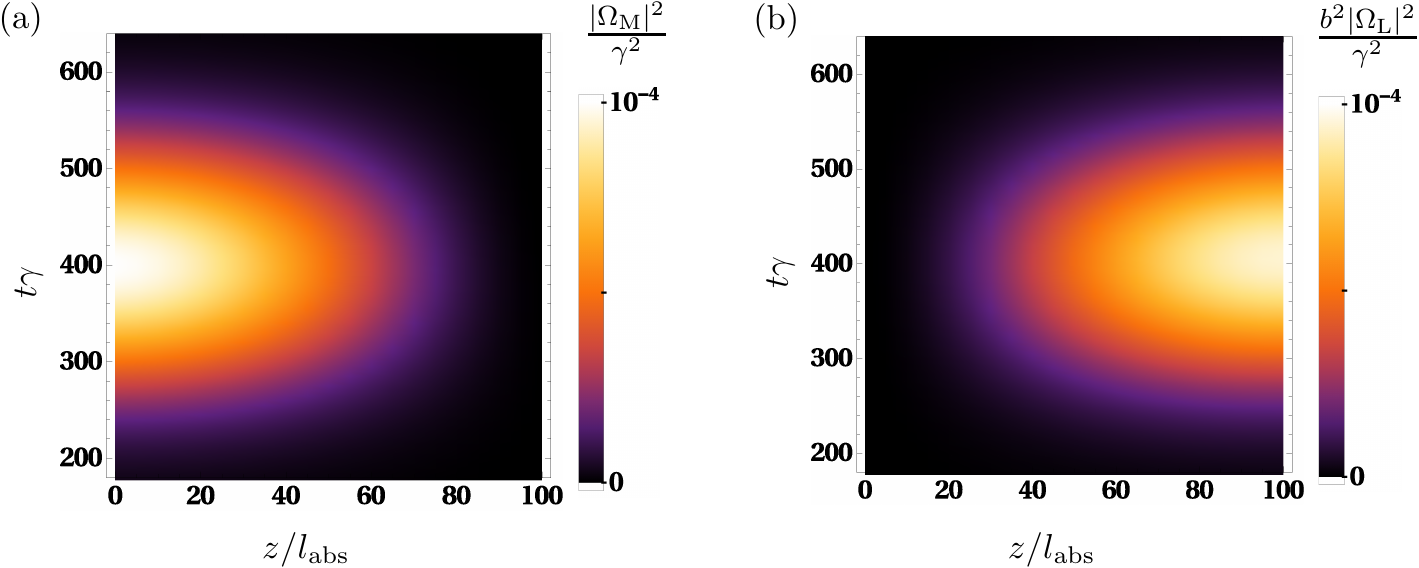}
\end{center}
\caption{\label{fig4}
Frequency conversion of pulsed fields.
(a) Density plot of the incoming millimeter wave pulse with a Gaussian envelope. 
(b) Density plot of the outgoing optical pulse.
The parameters in (a) and (b) are the same as in Fig.~\ref{fig2}.
}
\end{figure*}
%%%%%%%%%%%%%%%%%%%% 
%

%
The spatial oscillations of optical and mm-wave  intensities 
according to Eq.~(\ref{matrix}) are shown in Fig.~\ref{fig2}. Our model is in excellent 
agreement with a full numerical solution of the Maxwell-Bloch equations. Note that the small deviations for 
large $z$ vanish if the approximations leading to  Eq.~(\ref{matrix}) are omitted. 
Complete MMOC occurs after a length $L_c =\pi/(2 k)$, and thus 
requires an optical depth $D_c=L_c/l_{\text{abs}}$  that 
is inversely proportional to $\varepsilon$ in Eq.~(\ref{eps}), $D_c=\pi/(2\varepsilon)$. 
Since the value of $\varepsilon$  can be  adjusted through the intensities and 
frequencies of the auxiliary fields, the condition for complete MMOC can be 
met for various densities and sizes of atomic gases. 
In the example in Fig.~\ref{fig2}, we find  $L_c = 100 l_{\text{abs}}$. 
The efficiency $F=e^{-2 \kappa L_c}$  for complete conversion can be expressed 
in terms of the optical depth $D_c$, 
\begin{align}
 F(D_c) = \exp\left[-\frac{\pi^2}{2D_c}\right]\exp\left[ - 2 
\varepsilon_{\Gamma}D_c\right]\,,
 \label{efficiency}
\end{align}
and $F(D_c)$ is shown in Fig.~\ref{fig3} for three different values of $\varepsilon_{\Gamma}$. 
The maximum efficiency  $F_{\text{max}}=\exp(-2\pi\sqrt{\varepsilon_{\Gamma}})$ is attained at 
an optical depth  $D_c^{\text{max}}=\pi/(2\sqrt{\varepsilon_{\Gamma}})$ and tends to 
unity for $\varepsilon_{\Gamma}\rightarrow 0$. Since $\varepsilon_{\Gamma}\propto\Gamma$, 
efficiencies close to unity  are only possible because of the slow radiative  decay  rate $\Gamma$  of the Rydberg levels 
$\ket{3}$, $\ket{4}$ and $\ket{5}$. 
$\Gamma$  decreases   with increasing $n$ as $\Gamma\propto n^{-3}$~\cite{beterov:09} and is thus typically several 
orders of magnitude smaller than the decay rate $\gamma$ of the low-lying states $\ket{2}$ and $\ket{6}$. 
The efficiency for complete MMOC for the parameters in Fig.~\ref{fig2} is $F\approx 92.1\%$. 

Note that our definition of the  efficiency is based on photon fluxes and not intensities as required for a coherent conversion scheme 
that conserves the total  photon flux. In order to see this, we consider 
the perfectly coherent conversion of an optical field to a millimeter wave with $F(D_c)=1$.  According to Eq.~(\ref{matrix}), we 
obtain $|\Omega_{\text{M}}(L_c)|^2=b^2|\Omega_{\text{L}}(0)|^2$. With the definition of $b$ in Eq.~(\ref{bval}) and the definition of 
the Rabi frequencies we get
\begin{align}
 |\mc{E}_{\text{M}}(L_c)|^2=\frac{\omega_{\text{M}}}{\omega_{\text{L}}}|\mc{E}_{\text{L}}(0)|^2\,,
\end{align}
where $\mc{E}_{\text{M}}$ and $\mc{E}_{\text{M}}$ are the electric field amplitudes of the mm-wave and optical fields, respectively. 
The ratio of the intensities ($I\propto |\mc{E}|^2$) is thus $I_{\text{M}}^{\text{out}}/I_{\text{L}}^{\text{in}}= \omega_{\text{M}}/\omega_{\text{L}}$, as it should be. 
Similarly, we obtain $I_{\text{L}}^{\text{out}}/I_{\text{M}}^{\text{in}}=\omega_{\text{L}}/\omega_{\text{M}}$ for the conversion of mm-waves into optical fields. 

Next we consider the conversion of pulsed fields. The derivation of Eq.~(\ref{matrix})  
shows that our scheme is not mode-selective and  works for broadband pulses. 
The only requirement is that the atomic dynamics remains in the adiabatic regime, 
which holds if the  bandwidth $\delta\nu$ of the input pulse is 
smaller than all detunings $\Delta_k$ and the Rabi frequencies  $\Omega_{\text{R}}$, $\Omega_{\text{C}}$ and $\Omega_{\text{A}}$  (see Sec.~\ref{solution}).
In order to demonstrate this, we present numerical solutions of the Maxwell-Bloch equations  for 
a mm-wave input pulse as shown in Fig.~\ref{fig4}. 
The intensity of a mm-wave input pulse with Gaussian envelope is shown in 
Fig.~\ref{fig4}(a),  and the corresponding optical output field is shown in Fig.~\ref{fig4}(b). 
The input pulse has a bandwidth on the order of $\Delta\nu\approx 2\pi\times 80$~\text{kHz} and is converted without distortion of its shape. 
We thus find that the bandwidth of our conversion scheme is at least $\sim80$~\text{kHz} for the chosen parameters. 
This bandwidth can be significantly  increased  by increasing the detunings and Rabi frequencies of the auxiliary fields. 
Finally, we note that the conversion of optical pulses to mm-waves works equally well. 
\subsection{Rubidium parameters\label{real}}
%
%
%%%%%%%%%%%%%%%%%%%%%%
\begin{figure*}[t!]
\begin{center}
\includegraphics[width=12cm]{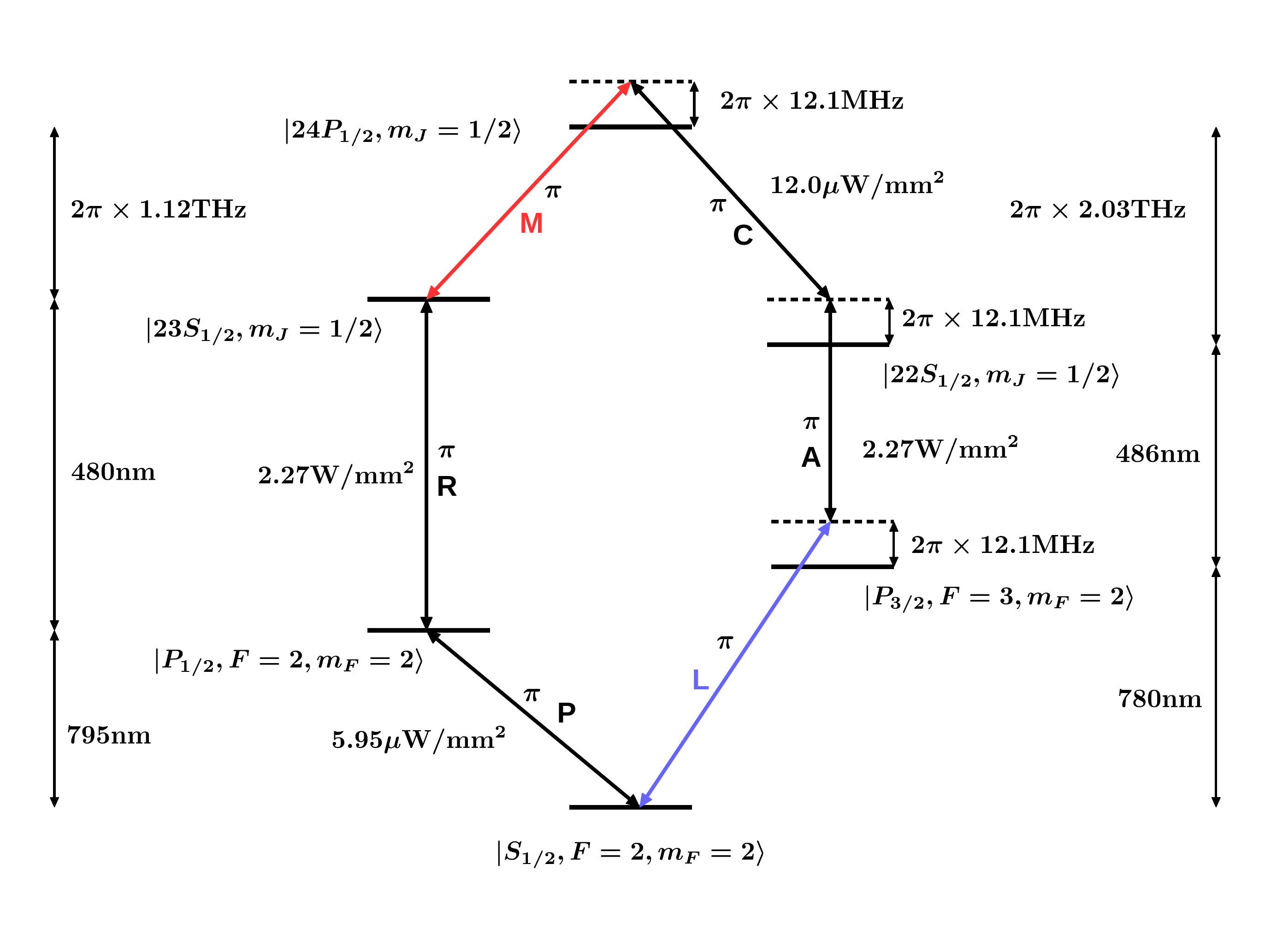}
\end{center}
\caption{\label{fig5}
Realisation of the level scheme in Fig.~\ref{fig1}  based on transitions in ${}^{87}$Rb.  
All quantum numbers of the employed states as well as intensities, polarisations and detunings are indicated. 
Note that energy spacings are not to scale. The intensities and detunings correspond to the parameters in Fig.~\ref{fig2}.
The decay rate $\gamma=2\pi\times 6.1 \text{MHz}$ corresponds to the $\text{D}_2$ line. We set the decay rate $\Gamma$ of all Rydberg states equal to the decay 
rate of the $\ket{23S_{1/2}}$ state at $T=300\text{K}$, which is faster than the decay rate of the $\ket{24P_{1/2}}$ state. We find~\cite{beterov:09} $\Gamma/\gamma= 1/285$ 
and the ratio of the coupling constants is   $b=\sqrt{0.72}$. 
}
\end{figure*}
%%%%%%%%%%%%%%%%%%%%%
%
%
Here we discuss one possible realisation of our scheme based on an ensemble of  ${}^{\text{87}}$Rb atoms. 
The atomic level scheme is shown in Fig.~\ref{fig5}, where
the optical field L couples to the $\text{D}_2$ line, and the auxiliary P field couples to the $\text{D}_1$ line. 
The transition dipole matrix elements for the optical transitions can be found in~\cite{steck:dl}, and 
for transitions between Rydberg states we follow the approach described in~\cite{walker:08}. 
The intensities of the auxiliary fields are chosen such that they correspond to the Rabi frequencies in Fig.~\ref{fig2}, and the 
values of the detuning parameters in Figs.~\ref{fig5} and~\ref{fig2} are also equivalent. 
Next we show that Rydberg-Rydberg interactions are negligible for the level scheme in Fig.~\ref{fig5} and for  
an atomic density of $\mc{N}=2\times10^{17}\text{m}^{-3}$. First we note that  the Rydberg blockade 
radius is $R_{\text{b}}\approx0.63\mu\text{m}$ for the parameters of Fig.~\ref{fig5}. This is significantly smaller than the mean distance between atoms, and hence 
the density of Rydberg atoms is simply given by $\mc{N}_{\text{Ry}}\approx \rho_{33}\mc{N}$~\cite{petrosyan:13,gaerttner:14}. 
By carrying out the average in Eq.~(\ref{averM}), we find that the matrix $\widetilde{\mc{M}}$ leads to the same conversion efficiency as  $\mc{M}$, i.e., 
there is no notable difference between the curves in Fig.~\ref{fig2} generated by $\mc{M}$ and the corresponding curves produced with $\widetilde{\mc{M}}$. 
On the other hand, if we choose $\ket{3}=\ket{24S_{1/2},m_J=1/2}$ instead of $\ket{3}=\ket{23S_{1/2},m_J=1/2}$,  
the conversion efficiency drops to $61\%$. 

In order to obtain more insight into these results, we consider the distance $R_{90}$ where $90\%$ of all Rydberg atom pairs will have a larger separation than $R_{90}$,
\begin{align}
\int_{R>R_{90}} \!\!\!\!\!\!\!\!\!\!\!\!\!\!\!\text{d}^3\!\mf{R}\;\,w(\mf{R}) = 0.9\,. 
\end{align}
Our parameters give $\mc{N}_{\text{Ry}}\approx 4.4\times10^{15}\text{m}^{-3}$ and thus 
$R_{90}\approx 1.79\mu\text{m}$. The van der Waals shift between two atoms in state $\ket{3}$ 
and separated by $R_{90}$ is~\cite{singer:05} 
\begin{align}
\Delta_{\text{vdW}}=-\frac{C_6}{R_{90}^6} \approx 2\pi\times 24.5 \,\text{kHz}. 
\end{align}
This is much smaller than all detuning parameters and Rabi frequencies entering the matrix $\mc{M}$. Since the frequency shifts for $90\%$ of all 
atoms are even smaller, averaging over all nearest neighbour distances does not change the matrix $\mc{M}$. 
Similarly, the dipole-dipole shifts in Eq.~(\ref{dd}) with $\ket{3}=\ket{24S_{1/2},m_J=1/2}$ are on the order of 
\begin{align}
  \Delta_{\text{DD}} \propto \frac{1}{4\pi\varepsilon_0\hbar}\frac{|\bra{3}\mf{\hat{d}}\ket{4}|^2}{R_{90}^3}\approx 2\pi\times 62.6 \,\text{kHz}, 
 \label{nsnp}
\end{align}
which is also small  compared to the detuning parameters and Rabi frequencies of the auxiliary fields. 
On the other hand, $\Delta_{\text{DD}}$ increases by a factor of 100 by choosing $\ket{3}=\ket{24S_{1/2},m_J=1/2}$ instead of $\ket{3}=\ket{23S_{1/2},m_J=1/2}$. 
This explains why the conversion efficiency drops significantly 
by using the strong $\ket{ns}\leftrightarrow\ket{np}$ transition instead of $\ket{(n-1)s}\leftrightarrow\ket{np}$ as in Fig.~\ref{fig5}. 

The absorption length for the field L is $l_{\text{abs}}=5.1 \times 10^{-2}\text{mm}$ for our parameters. 
Since full conversion requires an optical depth of $\sim$100, the length of the medium needs to be $L_c\approx 5.1\text{mm}$. 
These parameters are experimentally achievable. For example, much higher optical depths $\sim$1000 have been reported~\cite{hsiao:14,sparkes:13}, 
and the atomic cloud size considered here is similar to the dimensions of the experiment in~\cite{sparkes:13}, where  cold Rb atoms were trapped 
in a cylindrical  geometry of length $4.6\,\text{mm}$ and width $0.45\,\text{mm}$. 
\subsection{Physical implementation \label{imp}}
%
%
%%%%%%%%%%%%%%%%%%%%%%
\begin{figure*}[t!]
\begin{center}
\includegraphics[width=13cm]{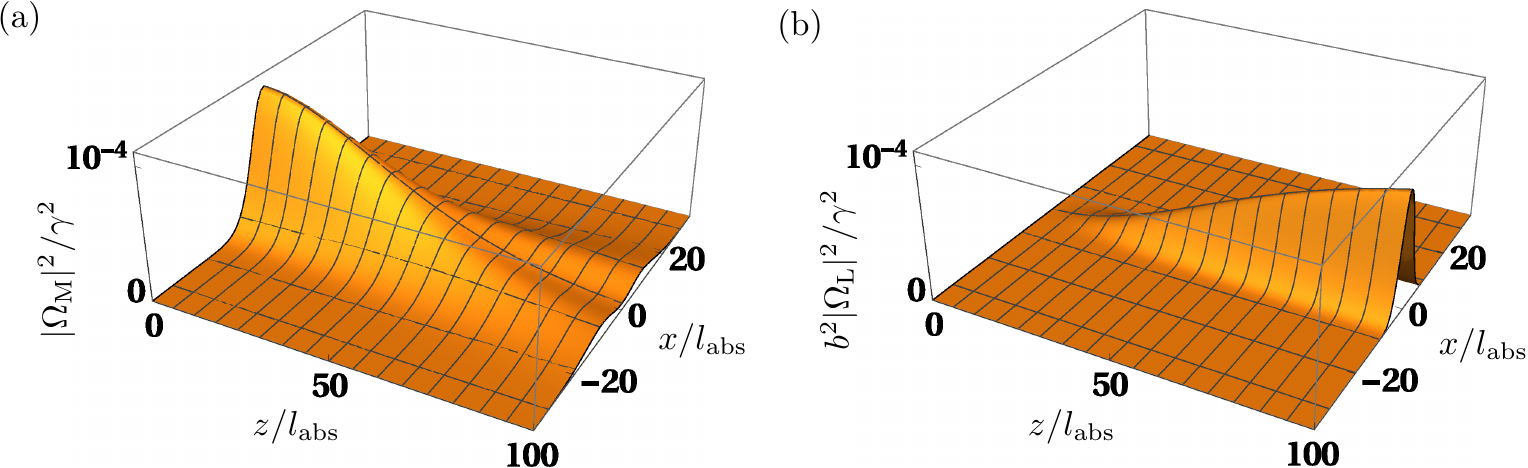}
\end{center}
\caption{\label{fig6}
(a) Spatial intensity profile $|\Omega_{\text{M}}|^2$ of an incident mm-wave field in the $x$-$z$ plane. 
The beam profile of the mm-wave at $z=0$ is given by Eq.~(\ref{FieldProfile}).
(b) Spatial intensity profile $|\Omega_{\text{L}}|^2$ of the resulting generated optical field in the $x$-$z$ plane. 
The fields in (a) and (b) are cylindrically symmetric, and the parameters  are $\sigma_c = 8 l_{\text{abs}}\approx 413\,\mu\text{m}$, 
$\sigma_{\text{M}}=1.9 \lambda_{\text{M}}\approx 509\,\mu\text{m}$, $\mc{N}^{(0)}=2\times 10^{17}\,\text{m}^{-3}$ and $\Omega_{\text{M}}^{(0)}=10^{-4}\gamma$. 
All other parameters are specified in Sec.~\ref{real}. 
}
\end{figure*}
%%%%%%%%%%%%%%%%%%%% 
%
We first consider the setup in Fig.~\ref{fig1}(a) where the mm-wave field is focussed into the atomic ensemble by  
lenses. We assume that the focal spot is at $z=0$ such that the mm-wave beam profile is 
\begin{align}
 \Omega_{\text{M}}(z=0,r)=\Omega_{\text{M}}^{(0)} e^{-r^2/\sigma_{\text{M}}^2},
 \label{FieldProfile}
\end{align}
where $r=\sqrt{x^2+y^2}$ is the radial coordinate in the $x$-$y$ plane, $\sigma_{\text{M}}$ is the beam waist and $\Omega_{\text{M}}^{(0)}$ is the peak Rabi frequency at the center of the beam. 
We model the transverse density profile of the atom cloud by a Gaussian with peak density $\mc{N}^{(0)}$ and width $\sigma_c$,
\begin{align}
 \mc{N}(r) = \mc{N}^{(0)}e^{-2 r^2/\sigma_c^2}, 
 \label{DensityProfile}
\end{align}
In order to calculate the conversion efficiency, 
we find the stationary solution of  Eq.~(\ref{maxS1d}) with the boundary condition in Eq.~(\ref{FieldProfile}), the density profile in Eq.~(\ref{DensityProfile}) 
and with the  analytical expression for the atomic coherences in Eq.~(\ref{psol2}). 
The result for the parameters specified in Sec.~\ref{real} is shown in Fig.~\ref{fig6}, where we consider 
a beam waist of $\sigma_{\text{M}}=1.9 \lambda_{\text{M}}$ and an atomic cloud with transverse size $\sigma_c \approx 413\,\mu\text{m}$. 
The intensity of the millimeter wave is shown in Fig.~\ref{fig6}(a) and decreases due to the conversion mechanism. In addition, it broadens 
slightly with increasing $z$ which can be understood as follows. For the given parameters the Rayleigh length $z_{\text{M}}=\pi \sigma_{\text{M}}^2/\lambda_{\text{M}}$ 
of the mm-wave is $z_{\text{M}} \approx11.3 \lambda_{\text{M}}$, which is about half the length of the medium. The broadening is thus caused by the strong 
focussing of the beam before it enters the atomic ensemble. Note that the Rayleigh length is  much larger than the wavelength $\lambda_{\text{M}}$, and hence the paraxial approximation is justified. 
The intensity of the optical wave is shown in Fig.~\ref{fig6}(a) and increases with increasing $z$. In order to quantify the conversion efficiency, 
we consider the total power of the incoming mm-wave and of the outgoing optical field, 
\begin{subequations}
 \label{power}
\begin{align}
 P_{\text{M}}^{\text{in}} & = \frac{  \pi\varepsilon_0 c \hbar^2}{ |\mf{d}_{34}|^2} \int\limits_0^{\infty} |\Omega_{\text{M}}(z=0,r)|^2 r \,\text{d}r \,, \\
 P_{\text{L}}^{\text{out}} & = \frac{ \pi\varepsilon_0 c \hbar^2}{ |\mf{d}_{16}|^2}  \int\limits_0^{\infty}  |\Omega_{\text{L}}(z=L_c,r)|^2 r \, \text{d}r \,,
\end{align}
\end{subequations}
where $\varepsilon_0$ is the dielectric constant. We then define the conversion efficiency by
\begin{align}
 F= \frac{\omega_{\text{M}}}{\omega_{\text{L}}} \frac{P_{\text{L}}^{\text{out}}}{P_{\text{M}}^{\text{in}}}, 
\end{align}
which is consistent with our definition of the conversion efficiency in Sec.~\ref{ideal}.  
We find $F\approx 26\%$ for the parameters in Fig.~\ref{fig6}, and this value can be further increased by increasing the transverse size of the atomic cloud. 
For example, for an atomic ensemble with transverse size $\sigma_c \approx 1\text{mm}$ we obtain $F\approx 61\%$. 
In addition, the conversion of optical fields to mm-waves works equally well. For a Gaussian optical beam of width $\sigma_{\text{L}}\approx 509\,\mu\text{m}$ and 
all other parameters as in Fig.~\ref{fig6}, we find $F\approx 24\%$. This value increases to $F\approx 72\%$ if the atomic cloud size is increased 
to $\sigma_c \approx 1\text{mm}$. However, increasing the 
transverse size of the atomic ensemble requires auxiliary fields with higher power in order to maintain the intensities shown in Fig.~\ref{fig5}.  
Next we discuss the implementation shown in Fig.~\ref{fig1}(b), where the mm-waves are confined by a waveguide and an elongated atomic cloud is trapped 
inside the waveguide core. This setting can be approximately 
described by the one-dimensional model in Eq.~(\ref{mateq}) if the ratio of the coupling constants in Eq.~(\ref{bval}) is replaced by
\begin{align}
 b^2_{\text{wg}}=\frac{A_{\text{L}}}{A_{\text{M}}} b^2\,, 
\end{align}
where $A_{\text{M}}$ is the effective area of the mm-wave guided mode,  and $A_{\text{L}}$ is the transverse size of 
the optical beam which is assumed to match the transverse density profile of the atoms~\cite{hafezi:09,kiffner:10b}. 
In principle, the setup in 
Fig.~\ref{fig1}(b) can thus be employed to interconvert mm-waves with longer wavelengths that cannot be focussed down to 
realistic dimensions of cold atom clouds. However, since $A_{\text{L}}/A_{\text{M}}\ll 1$, this results in smaller values of the parameter $\varepsilon\propto b_{\text{wg}}$ 
defined in Eq.~(\ref{pcoeff}) and thus in larger values of the optical depth required for complete conversion, $D_c=\pi/(2\varepsilon)$. 
In order to achieve the required optical depths, the atoms could be confined inside hollow core fibres
where extremely large optical depths have been observed~\cite{vorrath:10,blatt:14}. In addition, mm-waves can similarly be guided by photonic crystal fibres~\cite{han:02}. 
The strong coupling of atoms with mm-wave and optical fields required for efficient conversion might then be achievable by embedding a small hollow-core photonic crystal fibre 
into a larger mm-wave photonic crystal fibre.
\section{Summary \label{summary}}
We have shown that frequency mixing in Rydberg gases enables the coherent conversion 
between mm-wave and optical fields. 
Due to the numerous possibilities for choosing the $\ket{3}\leftrightarrow\ket{4}$ 
transition within the Rydberg manifold, our 
proposed MMOC scheme enables the conversion of various frequencies ranging from 
terahertz radiation to the microwave spectrum, that is 
for frequencies in the range $10-10,000$~GHz. 
The degree of conversion can be adjusted through the atomic density and the 
ancillary drive field intensities and frequencies. 
Conversion efficiencies  are limited by the lifetime of the Rydberg 
levels and dipole-dipole interactions between Rydberg atoms. 
Imperfections due to Rydberg interactions can be minimised in 
ensembles with low atomic densities and by the choice of the atomic states and parameters of 
the auxiliary fields. We have analysed a realistic implementation for  
the interconversion of terahertz and optical fields  
with an ensemble of trapped rubidium atoms, and find that the conversion efficiency can exceed 90\%. 

Efficient conversion requires a large spatial overlap between the mm-wave and optical fields, 
and we have discussed two possible scenarios how to achieve this. First, we 
have considered focussed terahertz beams and found that high conversion efficiencies 
are possible if the Rayleigh length of the beams is comparable to the length of the 
atomic cloud. Second, we investigated a setup where the mm-wave fields are 
transversally confined by a waveguide and the atoms are trapped inside the waveguide core. 
The optical depth required for 
complete conversion increases by $\sqrt{A_{\text{M}}/A_{\text{L}}}$  compared to the 
free-space implementation, where $A_{\text{M}}$ is the effective area of the mm-wave guided mode  
and $A_{\text{L}}$ is the transverse size of the atomic cloud. 
This waveguide setting enables high conversion efficiencies 
close to the theoretical limit set by the  lifetime of the Rydberg states and Rydberg interactions.
\begin{acknowledgments}
We thank the National Research Foundation and the Ministry of Education of
Singapore for support.
The research leading to these results has received
funding from the European Research Council under the European Union's Seventh
Framework Programme (FP7/2007-2013)/ERC Grant Agreement no. 319286 Q-MAC, and from the UK EPSRC through the standard grant EP/J000051/1, 
the programme grant EP/K034480/1 and through the Hub for Networked Quantum Information Technologies (NQIT). JN acknowledges support from a Royal Society University Research Fellowship. 
\end{acknowledgments}
\appendix
\section{Atomic coherences\label{solution}}
Here we derive the adiabatic solutions for the atomic coherences $\vroW_{43}$ and $\vroW_{61}$ in Eq.~(\ref{psol}). 
To this end, we assume that the  fields  $\Omega_{\text{M}}$ and $\Omega_{\text{L}}$  are sufficiently weak 
and expand  the atomic density operator as follows~\cite{kiffner:05,kiffner:12d}, 
\begin{align}
\vroW = \sum\limits_{k=0}^{\infty} \vroW^{(k)}, 
\label{expansion}
\end{align}
where $\vroW^{(k)}$ denotes the contribution to $\vroW$ in $k^{\text{th}}$ order in the  Hamiltonian 
\begin{align}
 H_{1}=-\hbar \left(\Omega_{\text{M}}A_{43} + \Omega_{\text{L}}A_{61}\right) +\text{H.c.}\,.
 \label{Hp}
\end{align}
The  solutions $\vroW^{(k)}$ can be obtained by re-writing the master 
equation~(\ref{master_eq}) as
\begin{align}
\mc{L} \vroW = \mc{L}_{0}\vroW -  \frac{\imag}{\hbar} [ H_{1} , \vroW ] \,,
\label{superM}
\end{align}
where the linear super-operator $\mc{L}_{0}$ is independent of $\Omega_{\text{M}}$ and $\Omega_{\text{L}}$. 
Inserting the expansion~(\ref{expansion}) into Eq.~(\ref{superM}) 
leads to the following set of coupled differential equations 
\begin{align}
&\dot{\vroW}^{(0)}=  \mc{L}_0 \vroW^{(0)} \,, \label{rho_0} \\
&\dot{\vroW}^{(k)} =  \mc{L}_0 \vroW^{(k)} -\frac{\imag}{\hbar}[H_{1} ,\vroW^{(k-1)}]\,,
\quad k>0\,.
\label{iterative}
\end{align}
Equation~(\ref{rho_0}) describes  the interaction of 
the atom with the  fields $\Omega_{\text{P}}$, $\Omega_{\text{R}}$, $\Omega_{\text{C}}$ and $\Omega_{\text{A}}$  to all orders and 
in the absence of $H_1$. 
Higher-order contributions to $\vroW$ can be obtained if Eq.~(\ref{iterative}) is solved iteratively. 
Equations~(\ref{rho_0}) and~(\ref{iterative}) must be solved under the
constraints $\text{Tr}(\vroW^{(0)})=1$ and $\text{Tr}(\vroW^{(k)})=0$ ($k>0$). 
The zeroth-order solution $\vroW^{(0)}$ is  the EIT dark state of the three-level ladder system $\ket{1}$, $\ket{2}$ and $\ket{3}$. 
For the special case $\Delta_3=0$ and if the small decay rate $\Gamma$ of state $\ket{3}$ is neglected, we find
\begin{subequations}
 \label{cpt}
\begin{align}
 \vroW^{(0)}_{11} & = \frac{|\Omega_{\text{R}}|^2}{|\Omega_{\text{P}}|^2+|\Omega_{\text{R}}|^2} \,, \\
 \vroW^{(0)}_{33} & = \frac{|\Omega_{\text{P}}|^2}{|\Omega_{\text{P}}|^2+|\Omega_{\text{R}}|^2} \,, \\
 \vroW^{(0)}_{13} & = - \frac{\Omega_{\text{P}}^*\Omega_{\text{R}}^*}{|\Omega_{\text{P}}|^2+|\Omega_{\text{R}}|^2} \,.
\end{align}
\end{subequations}
For $|\Omega_{\text{P}}|\ll|\Omega_{\text{R}}|$ the steady state is reached within several inverse decay times $1/\gamma$. 
In general, we obtain the zeroth-order solution $\vroW^{(0)}$ for $\Delta_3\not=0$ and substitute it in the first-order equation~(\ref{iterative}) 
with $k=1$. The formal solution of this differential equation is given by 
\begin{align}
 \vroW^{(1)}(t) = & \frac{\imag}{\hbar} \mc{L}_{0}^{-1} [ H_{1}(t) , \vroW^{(0)} ] \notag \\
 & -\frac{\imag}{\hbar} \mc{L}_{0}^{-1}\int\limits_0^t \text{d}t' e^{\mc{L}_{0}(t-t')} \partial_{t'}\left([ H_{1}(t') , \vroW^{(0)} ] \right),
 \label{adiabatic}
\end{align}
where we assumed $H_1(0)=0$. If $H_1(t)$ varies sufficiently slowly with time, the second term on the right-hand side
in Eq.~(\ref{adiabatic}) involving the time derivative of $H_1$ can be neglected. 
More precisely, this approximation is justified if the  bandwidth $\delta_{\nu}$ of the  pulses $\Omega_{\text{M}}$ and $\Omega_{\text{L}}$ is small as compared to the relevant differences 
between eigenfrequencies of  $H_0$. Through a numerical study we find that this condition is satisfied if all detunings $\Delta_k$   
($k\in\{4,5,6\})$ and the Rabi frequencies $\Omega_{\text{R}}$, $\Omega_{\text{C}}$ and $\Omega_{\text{A}}$ are large as compared to the bandwidth $\delta_{\nu}$. 
In general, the analytical expression  for the first-order density operator $\vroW$ is too bulky to display here. A special solution if the conditions in Eq.~(\ref{cond}) 
are met is given in Eq.~(\ref{psol2}).
\clearpage

\onecolumngrid

\vspace*{0.5cm}
\begin{center}
{\large\bf Supplemental Material for:\\[0.3cm]
Two-way interconversion of millimeter-wave and optical fields in Rydberg gases} 
\end{center}

\vspace*{0.5cm}
%\end{widetext}
\twocolumngrid

\centerline{\large\textbf{Detailed  model}}

\vspace*{0.5cm}
Here we derive the Maxwell-Bloch equations for our system from first principles. The electric field amplitude of the 
millimeter wave is $\mf{E}_{\text{M}}$,  and the optical field is denoted by $\mf{E}_{\text{L}}$.  
The other fields $\mf{E}_{\text{P}}$, $\mf{E}_{\text{R}}$ and  $\mf{E}_{\text{C}}$ are auxiliary fields facilitating the frequency conversion. 
We decompose  all electric fields as ($\text{X}\in\{\text{P},\text{R},\text{M},\text{C},\text{L}\}$)
\begin{align}
 \mf{E}_{\text{X}}= \mf{E}_{\text{X}}^{(+)}(\mf{r},t) + \text{c.c.}\, ,
\end{align}
where $\mf{E}_{\text{X}}^{(+)}$ is the positive frequency part of field X. 
The   positive frequency parts of  $\mf{E}_{\text{M}}$ and $\mf{E}_{\text{L}}$ are  defined as 
\begin{subequations}
 \label{MLfields}
\begin{align}
 \mf{E}_{\text{M}}^{(+)}(\mf{r},t)&= \mf{e}_{\text{M}}\mc{E}_{\text{M}}(\mf{r},t)~e^{\imag(\mf{k}_{\text{M}}\cdot\mf{r}-\omega_{\text{M}} t) } \,, 
 \label{Mfield} \\
\mf{E}_{\text{L}}^{(+)}(\mf{r},t)&=\mf{e}_{\text{L}}\mc{E}_{\text{L}}(\mf{r},t)~e^{\imag(\mf{k}_{\text{L}}\cdot\mf{r}-\omega_{\text{L}}t)} \,, 
\label{Lfield}
\end{align}
\end{subequations}
where $\mf{e}_{\text{M}}$ ($\mf{e}_{\text{L}}$) is the  unit polarisation vector,  
$\omega_{\text{M}}$ ($\omega_{\text{L}}$) is the central frequency, $\mf{k}_{\text{M}}$ ($\mf{k}_{\text{L}}$) is the wave vector 
and $\mc{E}_{\text{M}}$ ($\mc{E}_{\text{L}}$) is the envelope function of 
$\mf{E}_{\text{M}}$ ($\mf{E}_{\text{L}}$). 
The positive frequency parts of the auxiliary fields are given by 
\begin{subequations}
\begin{align}
\mf{E}_{\text{P}}^{(+)}(\mf{r},t) & = \mf{e}_{\text{P}}\mc{E}_{\text{P}}~e^{\imag(\mf{k}_{\text{P}}\cdot\mf{r}-\omega_{\text{P}} t) } \,, \\
\mf{E}_{\text{R}}^{(+)}(\mf{r},t) & = \mf{e}_{\text{R}}\mc{E}_{\text{R}}~e^{\imag(\mf{k}_{\text{R}}\cdot\mf{r}-\omega_{\text{R}}t) }  \,, \\
\mf{E}_{\text{C}}^{(+)}(\mf{r},t) & = \mf{e}_{\text{C}}\mc{E}_{\text{C}}~e^{\imag(\mf{k}_{\text{C}}\cdot\mf{r}-\omega_{\text{C}} t)} \,, \\
\mf{E}_{\text{A}}^{(+)}(\mf{r},t) & = \mf{e}_{\text{A}}\mc{E}_{\text{A}}~e^{\imag(\mf{k}_{\text{A}}\cdot\mf{r}-\omega_{\text{A}} t)} \,,
\end{align}
\end{subequations}
where $\mf{e}_{\text{X}}$, $\mc{E}_{\text{X}}$ and $\omega_{\text{X}}$ is the unit polarisation vector, envelope function and central frequency of 
field $\mf{E}_{\text{X}}$, respectively ($\text{X}\in\{\text{P},\text{R},\text{C},\text{A}\}$).
In order to simplify the notation, we introduce  atomic transition operators 
\begin{align}
A_{kl} = \ket{k}\bra{l},\quad A_{kl}^{\dagger}=A_{lk}\,.
\end{align}
In electric-dipole and  rotating-wave approximation, the  
Hamiltonian of each atom interacting with the six laser fields is 
\begin{align}
\tilde{H} =& \hbar\sum\limits_{k=2}^{5}\omega_{k}A_{kk} - \left( A_{21} \mf{d}_{21}\cdot\mf{E}_{\text{P}}^{(+)}  + A_{32} \mf{d}_{32}\cdot\mf{E}_{\text{R}}^{(+)} 
\right. \label{Hschroed} \notag  \\
&   + A_{43} \mf{d}_{43}\cdot\mf{E}_{\text{M}}^{(+)} + A_{45} \mf{d}_{45}\cdot\mf{E}_{\text{C}}^{(+)}  \notag\\
& \left. + A_{56} \mf{d}_{56}\cdot\mf{E}_{\text{A}}^{(+)}  + A_{61} \mf{d}_{61}\cdot\mf{E}_{\text{L}}^{(+)} \,+\,\text{H.c.}\right)\,,
\end{align}
where $\hbar\omega_{k}$ denotes the energy of state $\ket{k}$ with respect to the energy of level $\ket{1}$. 
The  matrix element of the electric dipole moment operator $\mf{\hat{d}}$ on the transition transition $\ket{k}\leftrightarrow\ket{l}$ is defined as 
\begin{align}
 \mf{d}_{kl}=\bra{k}\mf{\hat{d}}\ket{l}\,. 
\end{align}
We model the  time  evolution of the atomic system by a  master equation  
for the reduced density operator $\vroS$, 
\begin{align}
\partial_t \vroS & = - \frac{\imag}{\hbar} [ \tilde{H} , \vroS ] +\mc{L}_{\gamma}\vroS\,.
\label{master_eq_S}
\end{align} 
%%%%%%%%%%%%%%%%%%%%%%%%%%%%%%%%%%%%%%%%%%%%%%%%%%%%%%%%%%%%%%%%%
% Spontaneous emission
%%%%%%%%%%%%%%%%%%%%%%%%%%%%%%%%%%%%%%%%%%%%%%%%%%%%%%%%%%%%%%%%%
The last term in Eq.~(\ref{master_eq_S}) describes  spontaneous emission and 
is given by 
\begin{align}
\mc{L}_{\gamma}\vroS  = &-  \frac{\gamma}{2}
\left( A_{12}^{\dagger}A_{12}   \vroS  + \vroS A_{12}^{\dagger}A_{12}   - 2 A_{12}\vroS A_{12}^{\dagger} \right)  \notag \\
      &     -  \frac{\Gamma}{2}
\left( A_{23}^{\dagger}A_{23}   \vroS  + \vroS A_{23}^{\dagger}A_{23}   - 2 A_{23}\vroS A_{23}^{\dagger} \right)\,, \notag \\
  &   -  \frac{\Gamma}{2}
\left( A_{34}^{\dagger}A_{34}   \vroS  + \vroS A_{34}^{\dagger}A_{34}   - 2 A_{34}\vroS A_{34}^{\dagger} \right) \,, \notag \\
 &  -  \frac{\Gamma}{2}
\left( A_{54}^{\dagger}A_{54}   \vroS  + \vroS A_{54}^{\dagger}A_{54}   - 2 A_{54}\vroS A_{54}^{\dagger} \right) \,, \notag \\
 &  -  \frac{\Gamma}{2}
\left( A_{65}^{\dagger}A_{65}   \vroS  + \vroS A_{65}^{\dagger}A_{65}   - 2 A_{65}\vroS A_{65}^{\dagger} \right) \,. \notag \\
 &  -  \frac{\gamma}{2}
\left( A_{16}^{\dagger}A_{16}   \vroS  + \vroS A_{16}^{\dagger}A_{16}   - 2 A_{16}\vroS A_{16}^{\dagger} \right) \,. \notag 
\end{align}
While  the ground states $\ket{1}$ is assumed to be (meta-) stable, the states $\ket{2}$, $\ket{3}$, $\ket{4}$ and $\ket{5}$ 
decay through spontaneous emission. The decay rate $\gamma$ is the full decay rate of states $\ket{2}$ and $\ket{5}$, and 
$\Gamma$ is the decay rate on the Rydberg transitions.  In our scheme,  $\Gamma$ is much smaller than 
the decay rate  $\gamma$ of the low-lying electronic states. 
%%%%%%%%%%%%%%%%%%%%%%%%%%%%%%%%%%%%%%%%%%%%%%%%%%%%%%%%%%%%%%%%%%%%%
In order to remove the fast oscillating terms in Eq.~(\ref{master_eq_S}), we transform the latter equation  into a rotating frame 
\begin{align} 
W=\exp\left\{\imag[\right.& \omega_{\text{P}} A_{22}+(\omega_{\text{P}}+\omega_{\text{R}}) A_{33}\notag \\
&  +(\omega_{\text{P}}+\omega_{\text{R}}+\omega_{\text{M}}) A_{44} \notag \\
&  +(\omega_{\text{P}}+\omega_{\text{R}}+\omega_{\text{M}}-\omega_{\text{C}}) A_{55}  \notag \\
& \left. +(\omega_{\text{P}}+\omega_{\text{R}}+\omega_{\text{M}}-\omega_{\text{C}}-\omega_{\text{A}}) A_{66}]t \right\} \notag \\
\exp\left\{-\imag[\right.& \mf{k}_{\text{P}}\cdot\mf{r} A_{22}+(\mf{k}_{\text{P}}+\mf{k}_{\text{R}})\cdot\mf{r} A_{33} \notag \\
&  +(\mf{k}_{\text{P}}+\mf{k}_{\text{R}}+\mf{k}_{\text{M}})\cdot\mf{r} A_{44} \notag \\
&  +(\mf{k}_{\text{P}}+\mf{k}_{\text{R}}+\mf{k}_{\text{M}}-\mf{k}_{\text{C}})\cdot\mf{r} A_{55}\notag   \\
& \left. +(\mf{k}_{\text{P}}+\mf{k}_{\text{R}}+\mf{k}_{\text{M}}-\mf{k}_{\text{C}}-\mf{k}_{\text{A}})\cdot\mf{r} A_{66}] \right\} \,. \notag  
\end{align}
We assume that the central frequencies of all fields  are  resonant with the loop 
$\ket{1}\leftrightarrow\ket{2}\leftrightarrow\ket{3}\leftrightarrow\ket{4}\leftrightarrow\ket{5}\leftrightarrow\ket{6}\leftrightarrow\ket{1}$, 
\begin{align}
 \omega_{\text{P}}+\omega_{\text{R}}+\omega_{\text{M}}-\omega_{\text{C}}-\omega_{\text{A}}=\omega_{\text{L}}\,.
\end{align}
In addition, we impose the phase matching condition 
\begin{align}
 \mf{k}_{\text{P}}+\mf{k}_{\text{R}}+\mf{k}_{\text{M}}-\mf{k}_{\text{C}}-\mf{k}_{\text{A}}=\mf{k}_{\text{L}}\,.
\end{align}
The transformed density operator $\vroW=W\vroS W^{\dagger}$ obeys the master equation 
\begin{align}
\partial_t \vroW &= - \frac{\imag}{\hbar} [ H , \vroW ] +\mc{L}_{\gamma}\vroW\,, 
\end{align} 
and  the transformed Hamiltonian  $H$ is  
\begin{align}
H  = & - \hbar\sum\limits_{k=3}^{6} \Delta_{k} A_{kk}    \notag \\
& -\hbar \left( \Omega_{\text{P}}A_{21}+ \Omega_{\text{R}}A_{32}+\Omega_{\text{M}}A_{43} \right.\notag \\
&\qquad \left. + \Omega_{\text{C}}A_{45}+ \Omega_{\text{A}}A_{56} + \Omega_{\text{L}}A_{61}  \,+\,\text{H.c.}\right)\,.
\end{align}
In this equation,  $\Delta_k$ $k\in\{3,\ldots,6\}$ is a detuning defined as 
 \begin{subequations}
 \begin{align} 
\Delta_{3}=& \omega_{\text{P}}+\omega_{\text{R}} -\omega_{3}\,, \\
\Delta_{4}=& \omega_{\text{P}}+\omega_{\text{R}}+\omega_{\text{M}} -\omega_{4}\,, \\
\Delta_{5}=& \omega_{\text{P}}+\omega_{\text{R}}+\omega_{\text{M}} - \omega_{\text{C}} -\omega_{5}\,, \\
\Delta_{6}=& \omega_{\text{L}} -\omega_{6}\,.
\end{align}
\end{subequations}
The Rabi frequencies of the various fields are 
%%%%%%%%%%%%%%%%%%%%%%%%%%%%%%%%%%%%%%%%%%%%%%%%%%%%%%%%%%%%%%%%%%%%%%%
%                   Rabi's Frequency                                  %
%%%%%%%%%%%%%%%%%%%%%%%%%%%%%%%%%%%%%%%%%%%%%%%%%%%%%%%%%%%%%%%%%%%%%%%
 %
 \begin{subequations}
 \begin{align}
 \Omega_{\text{P}} & = \frac{\mf{d}_{21}\cdot\mf{e}_{\text{P}} }{\hbar}\mc{E}_{\text{P}} , \\
 \Omega_{\text{R}} & = \frac{\mf{d}_{32}\cdot\mf{e}_{\text{R}} }{\hbar}\mc{E}_{\text{R}} , \\
 \Omega_{\text{M}} & = \frac{\mf{d}_{43}\cdot\mf{e}_{\text{M}} }{\hbar}\mc{E}_{\text{M}} , \\
 \Omega_{\text{C}} & = \frac{\mf{d}_{45}\cdot\mf{e}_{\text{C}} }{\hbar}\mc{E}_{\text{C}} , \\
 \Omega_{\text{A}} & = \frac{\mf{d}_{45}\cdot\mf{e}_{\text{A}} }{\hbar}\mc{E}_{\text{A}} , \\
 \Omega_{\text{L}} & = \frac{\mf{d}_{51}\cdot\mf{e}_{\text{L}} }{\hbar}\mc{E}_{\text{L}} .
 \end{align}
 \label{Rabi}
 \end{subequations}
 %
%%%%%%%%%%%%%%%%%%%%%%%%%%%%%%%%%%%%%%%%%%%%%%%%%%%%%%%%%%%%%%%%%%%%%%%
%
Since $\Omega_{\text{M}}$ and $\Omega_{\text{L}}$ depend on 
position and time via the envelope functions $\mc{E}_{\text{M}}$ and $\mc{E}_{\text{L}}$, the 
density operator  $\vroW$ in the rotating frame is a  slowly varying  function  
of~$\mf{r}$ and~$t$.  

The propagation of the probe and control  fields  inside the 
medium is governed by Maxwell's equations. We only take into account $\mf{E}_{\text{M}}$ and $\mf{E}_{\text{L}}$ 
for the self-consistent Maxwell-Bloch equations. The P and R fields create coherent population trapping on the $\ket{1}\leftrightarrow\ket{2}\leftrightarrow\ket{3}$ 
transition such that the atoms are in a dark state for these fields. After a transient time, the P and R fields will thus not experience absorption or dispersion. 
Furthermore, the auxiliary fields C and A are detuned from resonance and couple to states that are virtually empty (see Supplementary Section `Analytical solution'). 
We can thus neglect their absorption and dispersion. 
The wave equation governing the propagation of the  electric field $\mf{E} = \mf{E}_{\text{M}} + \mf{E}_{\text{L}}$ is then given by 
\begin{align}
\left(\frac{1}{c^2}\partial_t^2 -\Delta\right)\mf{E}&= - \frac{1}{c^2\epsilon_0} \partial_t^2\mf{P}\,.
\label{waveEq}
\end{align}
The source term  on the right hand side of Eq.~(\ref{waveEq})  comprises  the macroscopic 
polarisation $\mf{P}$ induced by the external fields. We neglect atom-atom interactions such that 
$\mf{P}$ can be expressed in terms of the single-atom polarisation, 
%%%%%%%%%%%%%%%%%%%%%%%%%%%%%%%%%%%%%%%%%%%%%%%%%%%%%%%%%%%%%%%%%%%%%%%
%                       induced polarisation's                        %
%%%%%%%%%%%%%%%%%%%%%%%%%%%%%%%%%%%%%%%%%%%%%%%%%%%%%%%%%%%%%%%%%%%%%%%
%
\begin{align} 
\mf{P} &
= \mc{N}( \mf{d}_{34}\vroS_{43}  +\mf{d}_{16}\vroS_{61}  + \text{c.c.})\,. 
\label{pol}
 \end{align}
 %
%%%%%%%%%%%%%%%%%%%%%%%%%%%%%%%%%%%%%%%%%%%%%%%%%%%%%%%%%%%%%%%%%%%%%%%
In this equation, $\mc{N}$ is the  atomic density of the medium.  
Note that the coherences $\vroS_{43}$ and $\vroS_{61}$  in Eq.~(\ref{pol}) 
are related to the coherences of the density operator $\vroW$ in the rotating frame by 
\begin{align}
\vroS_{43} & = \vroW_{43}e^{\imag(\mf{k}_{\text{M}}\cdot\mf{r}-\omega_{\text{M}} t) }  \,,~~\vroS_{61}=\vroW_{61}e^{\imag(\mf{k}_{\text{L}}\cdot\mf{r}-\omega_{\text{L}} t)}\,.
\end{align}
If $\Omega_{\text{M}}$ and $\Omega_{\text{L}}$ propagate in $z$ direction, the wave equation~(\ref{waveEq}) can be cast into the form 
%
%%%%%%%%%%%%%%%%%%%%%%%%%%%%%%%%%%%%%%%%%%%%%%%%%%%%%%%%%%%%%%%%%%%%%%%
%                    Maxwell's Equations                               
%%%%%%%%%%%%%%%%%%%%%%%%%%%%%%%%%%%%%%%%%%%%%%%%%%%%%%%%%%%%%%%%%%%%%%%
\begin{subequations}
\label{max} 
\begin{align}
& \left[-\frac{\imag}{2k_{\text{M}}}\left(\Delta_{\perp}+\partial_z^2\right)+\frac{1}{c} \partial_t+ \mf{\hat{k}}_{\text{M}}\cdot\nabla\right]  \Omega_{\text{M}}  = \imag \eta_{\text{M}} \vroW_{43} \,,           \label{max1}\\
& \left[-\frac{\imag}{2k_{\text{L}}}\left(\Delta_{\perp}+\partial_z^2\right)+\frac{1}{c} \partial_t+ \mf{\hat{k}}_{\text{L}}\cdot\nabla\right]  \Omega_{\text{L}}  =  \imag \eta_{\text{L}} \vroW_{61}\,,   \label{max2}
 \end{align}
\end{subequations}
%%%%%%%%%%%%%%%%%%%%%%%%%%%%%%%%%%%%%%%%%%%%%%%%%%%%%%%%%%%%%%%%%%%%%%%
%
 where the coupling constants $\eta_{\text{M}}$ and $\eta_{\text{L}}$ are given by 
 \begin{subequations}
 \label{SlowMax}
 \begin{align}
 \eta_{\text{M}} & =\frac{\mc{N} |\mf{d}_{43}|^2}{2\hbar\epsilon_0 c} \omega_{\text{M}} \,, \\
 \eta_{\text{L}} & =\frac{\mc{N} |\mf{d}_{61}|^2}{2\hbar\epsilon_0 c} \omega_{\text{L}} \,, 
 \end{align}
 \end{subequations}
 and $c$ is the speed of light. 
 %
%
% Self-consistency
%
In the paraxial approximation, it is assumed that  the envelopes change slowly with $z$ as compared to the the wavelength of the fields, 
\begin{align}
 |\partial_z^2\Omega|\ll |k\partial_z\Omega|. 
\end{align}
By neglecting the second derivatives $\partial_z^2$ in Eq.~(\ref{max}), we obtain Eq.~(6) of the manuscript. 
 
%
% optical depth
%
Next we derive the expression for the absorption length  employed in the main text. To this end, we consider  
that all fields except for $\Omega_{\text{L}}$ are zero and find the steady-state coherence on the 
$\ket{6}\leftrightarrow\ket{1}$ transition in first order in $\Omega_{\text{L}}$, 
\begin{align}
\vroW_{61} = -\frac{\Delta_{\text{L}}-\imag\gamma/2}{\Delta_{\text{L}}^2+\gamma^2/4}\Omega_{\text{L}}. 
\label{twolevelC}
\end{align}
Next we substitute   Eq.~(\ref{twolevelC}) into Eq.~(\ref{max2}) and solve for the stationary state with boundary 
condition $\Omega_{\text{L}}(z=0)=\Omega_{\text{L}}^{(0)}$, 
\begin{align}
\Omega_{\text{L}}(z) =\Omega_{\text{L}}^{(0)} \exp( -2 \eta_{\text{L}}/\gamma z). 
\end{align}
The intensity $I_{\text{M}}$ is proportional to $|\Omega_{\text{L}}|^2$ and hence we obtain
\begin{align}
 I_{\text{M}}(z)= I_{\text{M}}^{(0)} \exp(-z/l_{\text{abs}}) , 
\end{align}
where the absorption length is $l_{\text{abs}} = \gamma/(4\eta_{\text{L}})$. After propagating through a medium of length $L$, 
the intensity has thus reduced to 
\begin{align}
 I_{\text{M}}(L)= I_{\text{M}}^{(0)} \exp(-D) , 
\end{align}
where $D=L/l_{\text{abs}}$ is the optical depth. 
\end{document}